\documentclass[prb,twocolumn,showpacs]{revtex4}
\usepackage[latin1]{inputenc}
\usepackage{graphicx}
\usepackage{amssymb}
\usepackage{amsmath}
\usepackage{citesort}
\usepackage{xspace}
\usepackage{dcolumn}
\usepackage{bm}
\usepackage{times}

\newfont{\tensy}{cmsy10}

\newcommand{\ie}[0]{i.e.\@\xspace}
\newcommand{\eg}[0]{e.g.\@\xspace}
\newcommand{\etal}[0]{et al.\@\xspace}

\newcommand{\op}{\hat{p}}
\newcommand{\ox}{\hat{x}}
\newcommand{\on}{\hat{n}}
\newcommand{\UP}[0]{\uparrow}
\newcommand{\DO}[0]{\downarrow}

\newcommand{\om}[0]{\omega}
\newcommand{\Ep}{E_\text{P}}
\newcommand{\nag}{{\phantom{\dag}}}

\newcommand{\dtau}{\Delta\tau}
\newcommand{\lc}{\lambda_\text{c}}

\newcommand{\las}[0]{\langle}
\newcommand{\ras}[0]{\rangle}
\newcommand{\la}[0]{\left\las}
\newcommand{\ra}[0]{\right\ras}

\renewcommand{\tilde}[1]{\widetilde{#1}}

\begin{document}

%%%%%%%%%%%%%%%%%%%%%%%%%%%%%%%%%%%%%%%%%%%%%%%%%%%%%%%%%%%%%%%%%%%%%%%%%%%%%
%%%%%%%%%%%%%%%%%%%%%       TITLE & ABSTRACT          %%%%%%%%%%%%%%%%%%%%%%%
%%%%%%%%%%%%%%%%%%%%%%%%%%%%%%%%%%%%%%%%%%%%%%%%%%%%%%%%%%%%%%%%%%%%%%%%%%%%%
\title{Quantum Monte Carlo results for bipolaron stability in quantum dots}
\author{Martin Hohenadler}\email{mh507@cam.ac.uk}
\author{Peter B. Littlewood}
\affiliation{%
Theory of Condensed Matter, Cavendish Laboratory, University of Cambridge,
Cambridge, CB3 0HE, United Kingdom}

\begin{abstract}
  Bipolaron formation in a two-dimensional lattice with harmonic confinement,
  representing a simplified model for a quantum dot, is investigated by means
  of quantum Monte Carlo simulations. This method treats all interactions
  exactly and takes into account quantum lattice fluctuations.  Calculations
  of the bipolaron binding energy reveal that confinement opposes bipolaron
  formation for weak electron-phonon coupling, but abets a bound state at
  intermediate to strong coupling. Tuning the system from weak to strong
  confinement gives rise to a small reduction of the minimum Fr\"ohlich
  coupling parameter for the existence of a bound state.
\end{abstract}


% 71.38.-k : Polarons and electron-phonon interactions
% 71.38.Ht : Self-trapped or small polarons
% 71.38.Mx : Bipolarons 
% 61.46.-w : Nanoscale materials  (for electronic transport in nanoscale materials, see 73.63.-b)
% 63.22.+m : Phonons or vibrational states in low-dimensional structures and nanoscale materials 
% 73.21.La : Quantum dots 
% 02.70.Ss : Quantum Monte Carlo methods
 
\maketitle

\section{Introduction}

Continuous experimental improvements in the preparation and investigation of
semiconductor quantum dots in recent years have sparked a lot of interest in
extending our theoretical understanding of charge carriers confined in such
quasi-zero-dimensional systems.\cite{quantumdots} Apart from
potential technological applications, quantum dots with tunable properties
represent a playground to compare theory and experiment.

The importance of the lattice degrees of freedom in, \eg, polar semiconductors has
led to a large number of studies on polarons and bipolarons.\cite{AlMo95} These
quasiparticles correspond to bound states of a single carrier in a self-induced
lattice distortion or of two electrons in a shared virtual
phonon cloud. Bipolarons manifest themselves in, \eg, tunneling
experiments\cite{WaOrPh97} or transport through molecular
dots.\cite{AlBr03,KoRavO06} 

Whereas polaron formation in quantum dots is by now quite well understood
(see Refs.~\onlinecite{YiEr91,Sa96,PoFoDeBaKl99,ChMu01,HoFe06} and
references therein), conflicting results exist on the stability of
bipolarons.\cite{MuCh96,WaOrPh97,PoFoDeBaKl99,SeEr00,SeEr02_2,KrMuCh05,HoFe06}
These calculations are based on variational treatments, with several works
employing strong-coupling or adiabatic approximations. It is known from
studies of polaron and bipolaron formation that such methods are not able to
fully capture the relevant physics,\cite{AlMo95,Marsiglio95,HovdL05} and
their use hence represents a possible source of the contradictory
findings. The aim of this paper is to resolve the above issues by applying an unbiased quantum
Monte Carlo (QMC) method.\cite{HoFe06}

Previous works apply continuum models and the effective mass
approximation. Here we argue that a lattice model is essential for several reasons.
First, bipolaron physics involves self-trapping of carriers due to strong
interaction with the lattice, a process governed by lattice fluctuations on
the scale of the unit cell.\cite{Ra06}  Second, for intermediate to strong coupling,
bipolarons are rather small, so that any continuum description is
expected to break down.  Third, quantum dots studied experimentally often
contain only a relatively small number of unit cells.

Bipolaron formation in a quantum dot model with local interactions has
recently been studied numerically.\cite{HoFe06} The purpose of this work is
to extend these calculations to a more realistic model with long-range
interactions---similar to the continuum models employed by other
authors---and to obtain results for the bipolaron binding energy.

\section{Model}

The Hamiltonian considered here takes the form
\begin{eqnarray}\label{eq:hamiltonian}\nonumber
  H
  &=&
  -t \sum_{\las i,j\ras,\sigma} c^\dag_{i,\sigma} c^{\nag}_{j,\sigma}
  +
  \sum_{i,j} u_{ij} \on_{i,\UP}\on_{j,\DO}
  +
  K\sum_i |\bm{r}_i|^2 \on_i
  \\
  &&+
  \frac{\om_0}{2}\sum_i \left(\ox_i^2 + \op_i^2 \right)
  -\alpha\sum_{i,j} f_{j,i} \on_i\ox_j
  \,,
\end{eqnarray}
with the long-range Coulomb interaction
\begin{equation}\label{eq:u(r)}
  u_{ij} = 
  \begin{cases}
    U           &,\quad \bm{r}_i=\bm{r}_j\\
    U/|\bm{r}_i-\bm{r}_j|  &,\quad \bm{r}_i\neq\bm{r}_j
  \end{cases}
  \,,
\end{equation}
and the interaction between an electron at $\bm{r}_i$ and the oscillator at $\bm{r}_j$,
\begin{equation}\label{eq:f_ji}
  f_{j,i} = \frac{1}{(|\bm{r}_j-\bm{r}_i|^2+1)^{3/2}}
  \,.
\end{equation}
In Eq.~(\ref{eq:hamiltonian}), $c^\dag_{i,\sigma}$ creates an electron with
spin $\sigma$ at lattice site $i$ (located at $\bm{r}_i$), $\ox_i$ ($\op_i$) denotes
the displacement (momentum) of the harmonic oscillator at site $i$,
$\on_{i,\sigma}=c^\dag_{i,\sigma}c^\nag_{i,\sigma}$ and $\on_i=\sum_\sigma
\on_{i,\sigma}$.  The model parameters are the nearest-neighbor hopping
integral $t$, the Coulomb repulsion $U$, the confinement 
strength $K$, the (dispersionless) optical phonon frequency $\om_0$ and the
electron-phonon coupling constant $\alpha$. We analyze a
two-dimensional (2D) square lattice with two electrons of opposite spin and
periodic boundary conditions. 

For $|\bm{r}|\geq1$, the Coulomb interaction $u_{ij}$ in
Eq.~(\ref{eq:u(r)}) is of the same form as in previous studies of
bipolarons in quantum dots.\cite{MuCh96,PoFoDeBaKl99,SeEr00,KrMuCh05} An
important difference between previous work and our lattice model is that
$u_{ij}$ does not diverge for $\bm{r}_i=\bm{r}_j$. Instead, there is a
finite Hubbard $U>0$ for two electrons at the same site, which we believe to
be more appropriate for a discussion of small bipolaron states. 

The electron-phonon interaction [Eq.~(\ref{eq:f_ji})] can be regarded as a lattice version of
the Fr\"{o}hlich model.\cite{FrPeZi50,AlKo99,FeLoWe00} The Holstein-Hubbard
model\cite{HoFe06} is recovered by setting $u_{ij}=U\delta_{ij}$
and $f_{j,i}=\delta_{ij}$. We define\cite{AlKo99,FeLoWe00}
$\lambda = 2\tilde{\Ep}/W$ and $\tilde{\Ep}=\Ep\sum_j f_{j,0}^2,$
with the atomic-limit polaron binding energy of the Holstein-Hubbard model,
$\Ep=\alpha^2/2\om_0$, and the free bandwidth $W = 8t$. In the sequel,
all energies will be measured in units of $t$, and $N$ denotes the lattice
size in each dimension. We further set the lattice constant to 1. The 2D case models a
disk-shaped quantum dot with negligible thickness. The dependence on
dimensionality has been studied for the Holstein-Hubbard model,\cite{HoFe06}
and is weak for the Fr\"ohlich model.\cite{AlKo99}

The lattice Hamiltonian [Eq.~(\ref{eq:hamiltonian})] is appropriate in the study
of self-trapping, strong electron-phonon coupling or strong confinement. Compared to other
calculations of phonon effects in quantum dots in the framework of the
independent boson model,\cite{StZiCa00} we take into account the
finite size of the dot and hence the possibility of electronic hopping. This
is essential because (bi)polaron formation is determined by the balance
between kinetic, lattice, and Coulomb energies.

The harmonic (parabolic) confinement potential in
Eq.~(\ref{eq:hamiltonian}) is centered around site $(0,0)$ [the
lattice extends from $-(N-1)/2$ to $(N-1)/2$ in each dimension] and is
usually assumed to be a good approximation for real quantum
dots.\cite{quantumdots} The use of optical bulk phonons is sensible as 
bipolaron formation is dominated by the coupling to such branches,\cite{WaOrPh97} and the
details of the phonon spectrum are of minor importance.\cite{PoFoDeBaKl99}

The Hamiltonian [Eq.~(\ref{eq:hamiltonian})] contains the relevant terms to
describe bipolaron formation in quantum dots. Although more general cases can
also be treated with the present method,
our lattice model---more amenable to exact numerical treatments than the
continuum version---is chosen to be as similar as possible to the models
in existing work on bipolaron formation in order to resolve the conflicts
in available results. 

Finally, due to the simplicity of our model, we refrain from fitting our
results to experiment by tuning parameters. Instead, we chose the
adiabaticity ratio $\gamma=\om_0/t=0.1$ (\ie, in the experimentally important
adiabatic regime $\gamma\ll1$) and $U/t=4$ as in previous work,\cite{WeRoFe96,HovdL05,HoFe06}
and vary the electron-phonon coupling and the confinement strength. Note that within the
Fr\"ohlich model, $U$ and $\alpha$ are not independent
parameters.\cite{FrPeZi50}
 
\section{Method}

The worldline QMC method\cite{dRLa82,deRaLa86,HoFe06} can
be extended to the case of long-range electron-electron
interaction\cite{deRaLa86,Hague} and an electron-phonon coupling of the Fr\"ohlich
type.\cite{dRLa85,AlKo99} Alternative QMC schemes have
also been used to study bipolaron formation.\cite{Mac04,HovdL05}

Apart from the controlled (and small) Trotter error, no approximations are
made. In particular, the quantum-mechanical nature of the phonon degrees of
freedom is fully taken into account, and all interactions are treated on the
same footing. For the calculations, we have used a low temperature $\beta
t=t/(k_\text{B}T)=15$, a sufficiently small Trotter parameter $\dtau=0.05$,
and a linear lattice size $N=31$, ensuring small finite-size
effects. As pointed out before,\cite{HoFe06,HoLa07}
autocorrelations can be large and must therefore not be ignored.

The bipolaron binding energy is defined as
\begin{equation}\label{eq:EB}
  E_\text{B}
  =
  E(2)-2E(1)
  \,,
\end{equation}
where $E(N_\text{e})$ denotes the total energy of the system with
$N_\text{e}$ electrons.\footnote{Here we use the fermionic energy
$E^\text{f}=-\partial(\log Z^\text{f})/\partial\beta$ without the
free-phonon contribution, where $Z^\text{f}$ denotes the
fermionic part of the partition function.\cite{HoFe06}} Krishna \etal\cite{KrMuCh05}
introduced an additional estimated Coulomb correlation term of two unbound polarons in
a quantum dot, which generally leads to enhanced binding energies. Another
criterion for the stability of a bipolaron based on the relative distance of
the two electrons has also been suggested,\cite{SeEr02_2} yielding a broader
region of existence. Definition~(\ref{eq:EB}) has been used by most
previous authors, and permits comparison to the case $K=0$. Furthermore,
$E_\text{B}$ as defined by Eq.~(\ref{eq:EB}) has a direct physical meaning,
as it enters the Boltzmann factor that controls the average number of doubly
occupied quantum dots in an ensemble of dots.\cite{PoFoDeBaKl99} Note that we
use the same exact method to calculate $E(1)$ and $E(2)$.

We also measure the average electron-electron separation
\begin{equation}\label{eq:R}
  R = \la \sum_{i,j} (i-j)^2 \on_{i,\UP} \on_{j,\DO} \ra^{1/2}
  \,.
\end{equation}
For $\lambda=0$, $R$ gives an estimate of the size of the quantum dot,
whereas for $\lambda>0$ it may be regarded as a measure for the bipolaron
radius. For the calculation of other observables we refer to a previous
paper.\cite{HoFe06}

\section{Results}

A bipolaron is defined as a bound state of two electrons coupled to the
lattice.  The energy gain as compared to a system with two noninteracting
polarons originates in the additional potential energy from sharing a common
lattice polarization cloud. In the atomic limit $t=0$, the latter increases
quadratically with the number of electrons per site. This leads to a tendency
of the electrons to share the same region of space (\ie, to real-space
pairing), and gives rise to a bound state for any $\lambda>0$ if
  $U=0$, or for $\lambda>\lc(U)$ in the case $U>0$
where the electron-electron Coulomb repulsion has to be
overcome.\cite{Mac04,HovdL05}

\begin{figure}\centering
  \includegraphics[height=0.34\textwidth]{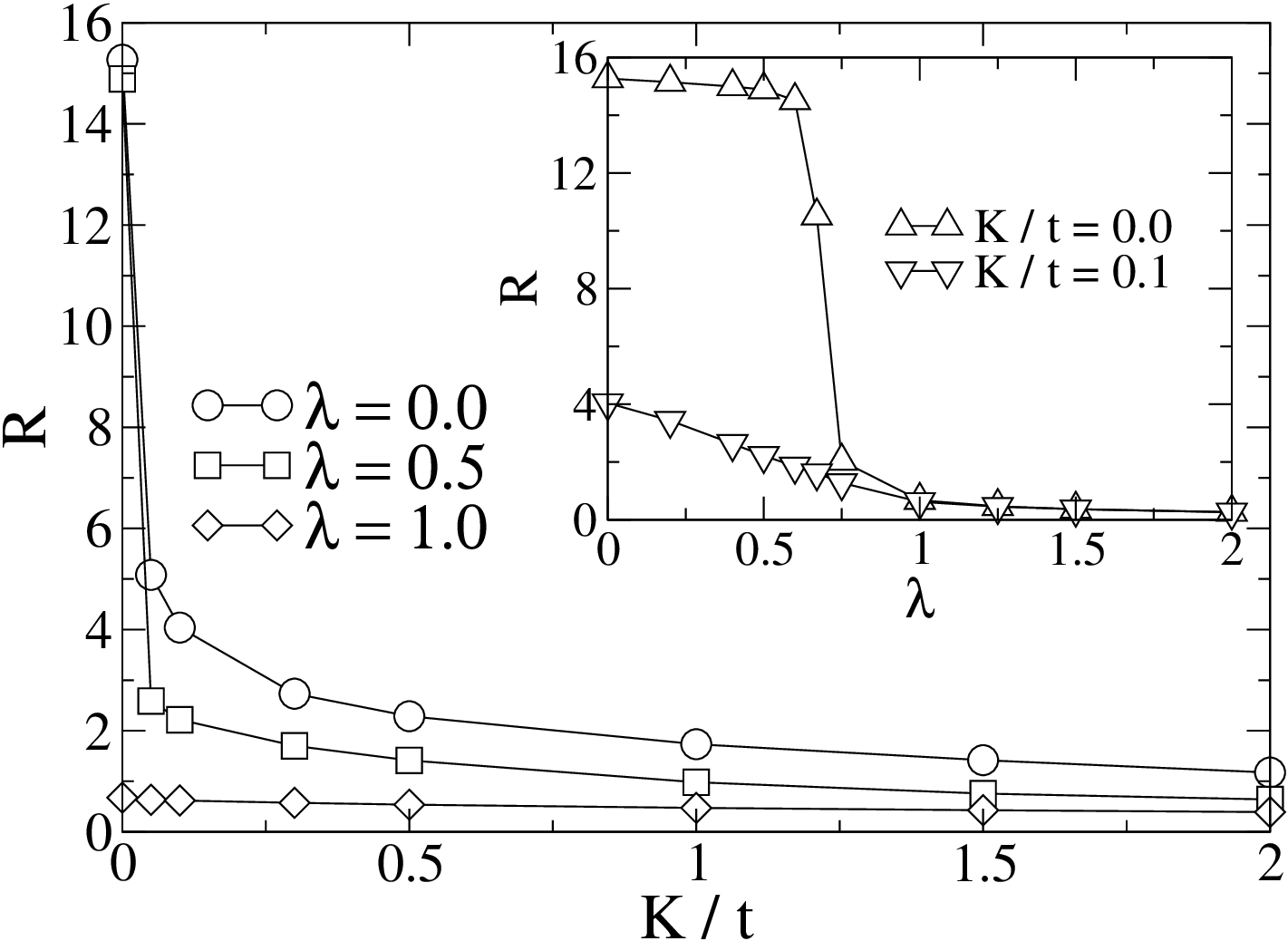}
  \caption{\label{fig:R_K} Average electron-electron distance $R$ as a function of
    confinement strength $K$ for different values of the electron-phonon coupling
    parameter $\lambda$. The inset shows $R$ as a function of $\lambda$ for
    two values of $K$. Here $\gamma=0.1$, $U/t=4$, $\beta t=15$
    ($\dtau=0.05$), and $N=31$. Lines are guides to the eye, and error bars
    are smaller than the symbols.}
\end{figure}

Depending on the model and the parameters, the average distance between the
particles can be larger than 1 (large bipolaron), about 1 (intersite
bipolaron) or less than 1 (small or onsite bipolaron).\cite{HovdL05} For $K=0$, the crossover from
a state with two unbound polarons to a bipolaron with increasing $\lambda$
can be detected from observables such as electron-electron correlation
functions or the ``radius'' $R$ defined by Eq.~(\ref{eq:R}) (see inset of
Fig.~\ref{fig:R_K}). However, in a confined system, the two-particle
wave function is ``squeezed'' even for $\lambda=0$,\cite{SeEr02_2,HoFe06} and the
binding energy is the only reliable indicator, with $E_\text{B}<0$ for a
bound state.

In some previous work on the continuum model, the strength of the confinement
potential was measured in terms of a confinement length. Therefore, we
start by analyzing the average electron-electron separation as a function of
$K$. Figure~\ref{fig:R_K} shows results for three different values of
$\lambda$. For $\lambda=0$ the distance rapidly decreases from $R\approx
N/2$ at $K=0$ to $R\approx2$--5 at finite $K$. For intermediate coupling
$\lambda=0.5$ the dependence is similar, but $R$ is
systematically smaller than for $\lambda=0$ due to the bipolaron
effect. Finally, for strong coupling $\lambda=1$, a small bipolaron is the
ground state even for $K=0$, so that confinement has very little
influence on $R$. The range of $R$ in
Fig.~\ref{fig:R_K} is comparable to the confinement lengths
studied by other authors,\cite{KrMuCh05} and the different behavior for
weak and strong couplings is consistent with continuum calculations.\cite{PoFoDeBaKl99}

The direct contribution of the confinement term in Eq.~(\ref{eq:hamiltonian}) 
to the total energy is linear in $N_\text{e}$, and therefore cancels
when calculating $E_\text{B}$. Consequently, any effect of $K$ on
$E_\text{B}$ is indirect, mediated by changes in the interaction energies
related to lattice distortions and Coulomb correlation.

\begin{figure}\centering
  \includegraphics[height=0.34\textwidth]{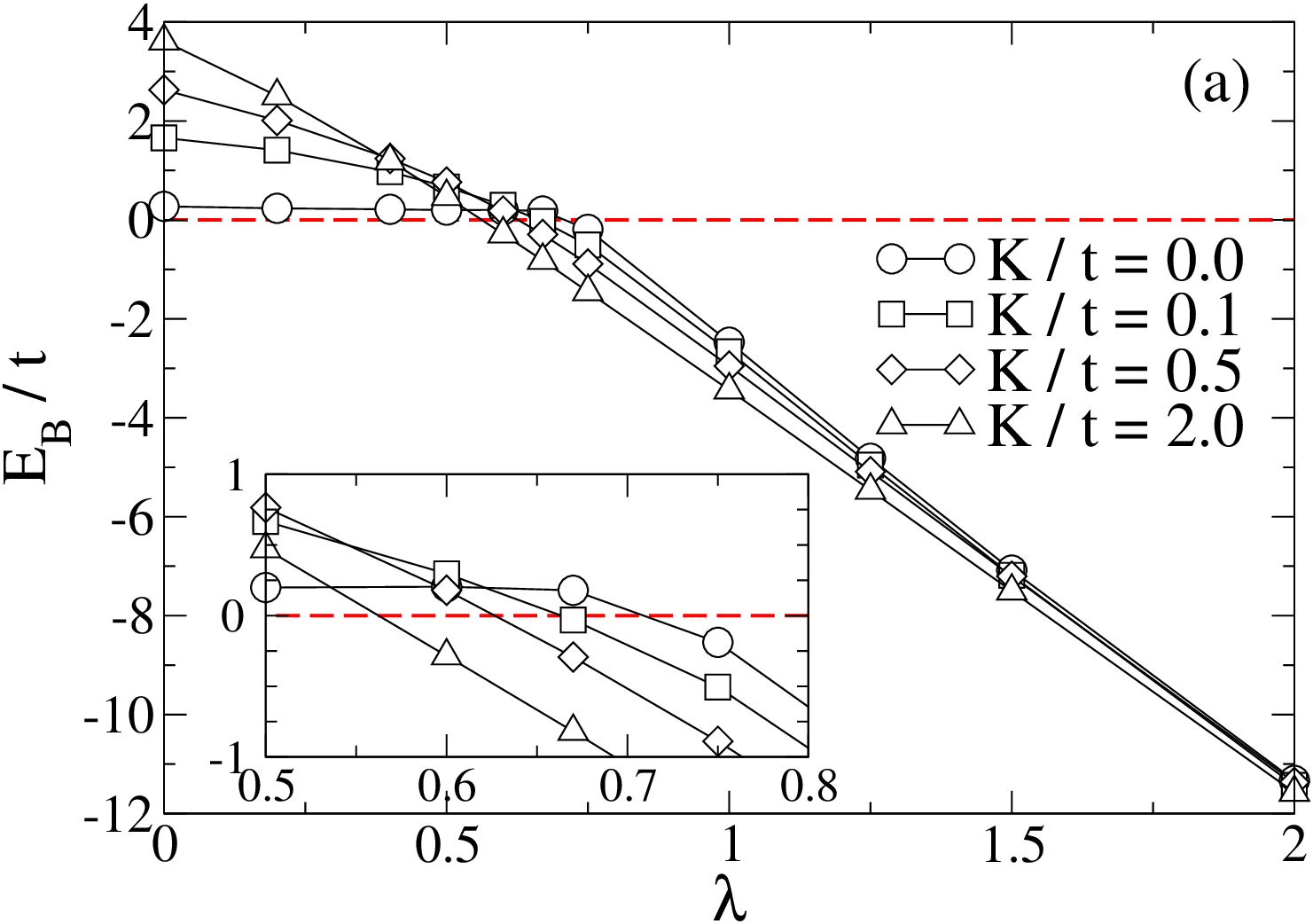}\\
  \hspace*{0.5em}
  \includegraphics[height=0.34\textwidth]{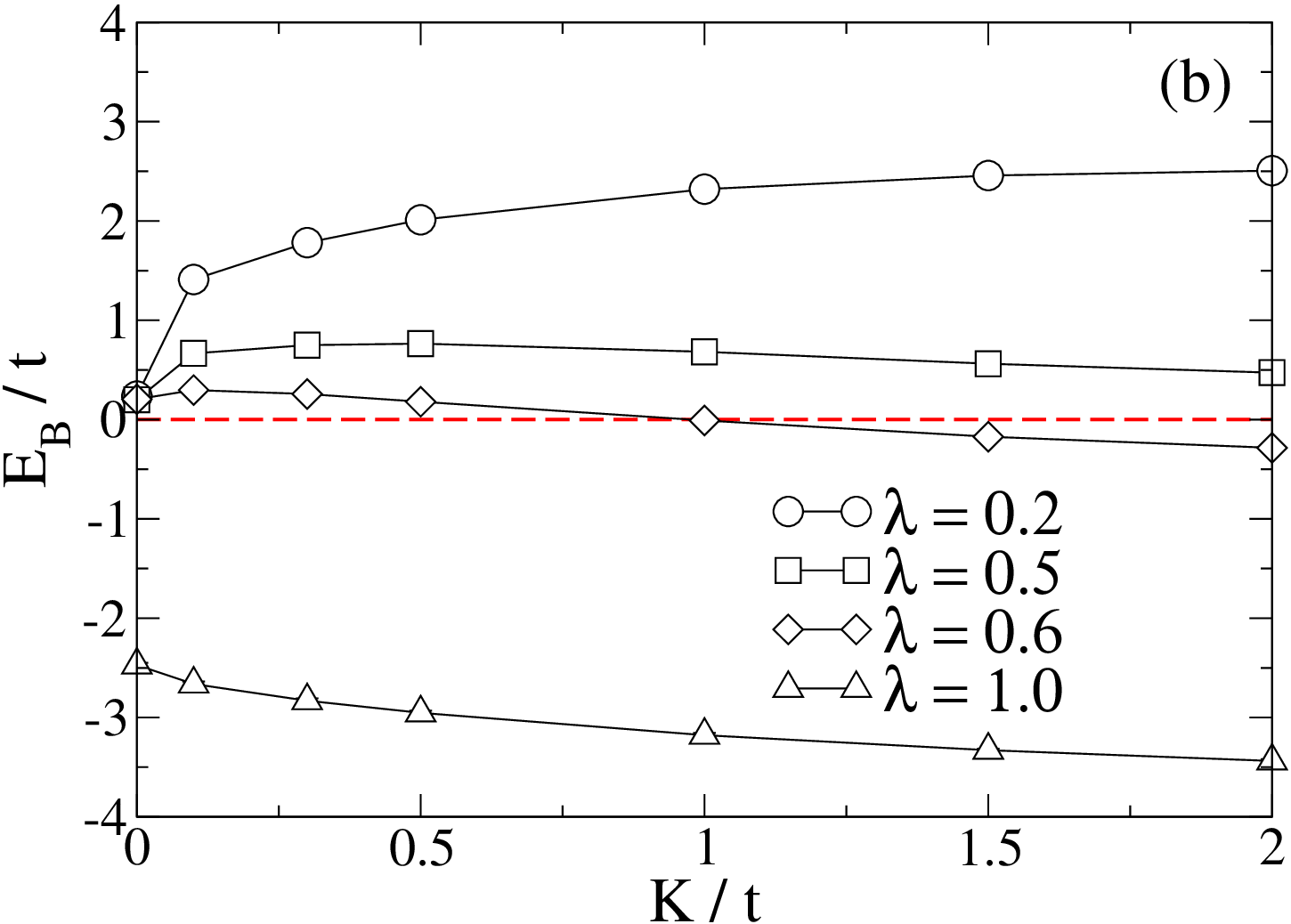}
  \caption{\label{fig:EB_lr} (Color online)
  Bipolaron binding energy (a)
  $E_\text{B}$ as a function of $\lambda$ for different $K$, and (b)
  $E_\text{B}$ as a function of $K$ for different $\lambda$. The dashed
  horizontal line indicates $E_\text{B}=0$, and the inset in (a) shows a
  closeup view.}
\end{figure}

In Fig.~\ref{fig:EB_lr}(a) we plot the bipolaron binding energy
$E_\text{B}$ as a function of $\lambda$ for
different values of $K$. As expected, all curves are strictly monotonic
decreasing with increasing electron-phonon coupling. The
binding energy is positive (unbound state) for $\lambda<\lc$, and negative
for $\lambda\gtrsim\lc$ (bound state or bipolaron). The inset in
Fig.~\ref{fig:EB_lr}(a) reveals a reduction of $\lc$ with increasing $K$ by
about 10\%, and the $K=0$ critical coupling is similar to that of the
Holstein-Hubbard model [cf. Fig.~7(a) in Ref.~\onlinecite{HoFe06}]. Of
course $\lc$ depends on $U$, with $\lc\to0$ as $U\to0$. For strong coupling,
all curves in Fig.~\ref{fig:EB_lr}(a) eventually collapse due to the
formation of a small-bipolaron state which is rather insensitive to confinement.

From Fig.~\ref{fig:EB_lr}(a), we can identify three different regimes, for which we plot
the binding energy as a function of $K$ in Fig.~\ref{fig:EB_lr}(b). 
In the regime $\lambda\ll\lc$, Coulomb interaction is stronger than the
phonon-mediated electron-electron attraction, so that we have two polarons
with $E_\text{B}>0$. Confinement enhances both interactions, and our
results reveal that the Coulomb energy becomes even more dominant,
\ie, $E_\text{B}$ increases with increasing $K$ for $K/t\lesssim1$. This is
particularly evident for the case $\lambda=0$ shown in
Fig.~\ref{fig:EB_lr}(a), where $E_\text{B}$ increases noticeably with $K$,
approaching the value of the onsite Coulomb repulsion $U/t=4$.

For $\lambda\gg\lc$, a bound state exists as a result of the lattice deformation energy
winning over the Coulomb repulsion, and confinement acts in favor of
bipolaron binding by further increasing $|E_\text{B}|$. The influence of
$K$ on the results is much weaker than that for $\lambda\ll\lc$ because the
physics is dominated by local correlations, as reflected by the small
bipolaron radius (see inset of Fig.~\ref{fig:R_K}). The weak dependence of
$\lc$ on $K$ may be attributed to the fact that confinement increases both
Coulomb repulsion as well as the phonon-mediated attractive interaction, and
thus leaves the effective interaction and $\lc$ almost unchanged.

In between these two limits, for intermediate $\lambda$, $E_\text{B}$ shows a
nonmonotonic behavior as a function of $K$. For intermediate coupling, 
we find a maximum in $E_\text{B}$ at small values of $K/t$, and a decrease
for an even stronger confinement. Whereas for $\lambda=0.5$ $E_\text{B}>0$ for
all $K$ shown, a bound state arises for a strong-enough confinement
$K/t\gtrsim1$ and $\lambda=0.6$.  In both cases, the variation of $E_\text{B}$ with $K$ is
rather small. For large $K$, the curves in Fig.~\ref{fig:EB_lr}(b) approach
the corresponding values of the effective onsite interaction $U-2\tilde{\Ep}$
which plays an important role in molecular quantum dots with only a single electronic
level.\cite{AlBr03,KoRavO06}

The main findings of previous calculations for the continuum Fr\"ohlich model
are as follows.\cite{SeEr02_2,PoFoDeBaKl99,HoFe06} There exists a minimal
$\lambda_\text{c}(U)$ below which a bound state is suppressed due to
Coulomb repulsion. Moreover, some authors argue that there is also
a critical confinement strength beyond which the diverging Coulomb repulsion
suppresses a bound state for any finite electron-phonon coupling
strength $\lambda<\infty$.\cite{MuCh96,SeEr00,KrMuCh05} No unanimous conclusion has been reached
concerning the effect of (weak to intermediate) confinement on the size of
the parameter region of existence for the bound bipolaron state.\cite{KrMuCh05}

Let us relate these findings to our results. Similar to previous work and
models without confinement, we find a critical coupling $\lc$ for the
formation of a bound state. As pointed out before, within our lattice
model, $\lc$ changes slightly as a function of the confinement
strength $K$, and scales approximately linearly with $U$.\cite{HoFe06}

Concerning the regime of strong confinement, our choice of a
finite Coulomb repulsion of two electrons with opposite spin located at the
same lattice site leads to substantially different physics. Even for
$K\to\infty$, a bound onsite bipolaron is formed if the effective onsite
interaction $U-2\tilde{\Ep}<0$. Finally, we observe that confinement opposes binding
for weak coupling $\lambda\ll\lc$, whereas it enhances binding for
intermediate to strong coupling $\lambda\gtrsim\lc$. Despite the associated
changes of the value of the binding energy, the phase diagram (\ie, the region of existence
of a bound state) is only weakly affected by the confinement potential. This
insensitivity is even more pronounced for the Fr\"ohlich parameter
$\alpha_\text{c}\propto\sqrt{\lc}$, which decreases with increasing $K/t\in[0,2]$ by
about 5\%.

Previous authors have used material-specific parameters to make predictions
for the existence of bipolarons in typical quantum dot
systems.\cite{KrMuCh05} As we believe that our model is too simple to make
quantitative statements, we restrict ourselves to a mostly qualitative discussion.
The (small) reduction of the critical coupling due to confinement suggests that the
dot size may determine whether bipolarons are stable or not. However,
according to our findings, this should only be relevant for materials with
intermediate electron-phonon coupling. In contrast, for weak
coupling, bipolaron formation is suppressed by Coulomb repulsion, whereas for
strong coupling a bipolaron ground state is stable regardless of the
confinement strength. Besides, as demonstrated in Fig.~\ref{fig:EB_lr}(b),
the value of the binding energy does have a noticeable dependence on the
confinement potential, at least for experimentally relevant weak to
intermediate values of the electron-phonon coupling. The dot size
may also change the phonon spectrum or electronic band structure, but such
effects have been neglected in our model.  

Finally, we can estimate the order of magnitude of the bipolaron binding
energy in real systems by assuming a typical bandwidth of 1 eV. From
Fig.~\ref{fig:EB_lr}, we conclude that for  plausible values of
$0.5\lesssim\lambda\lesssim1$, $E_\text{B}$ is a fraction of
an eV for $U=0.5$ eV, a value falling into the narrow-band regime $U\gg
t$. The binding energy increases with decreasing $U$, but the physics of bipolaron
formation remains qualitatively the same. Our values of $E_\text{B}$ are larger than
previous variational 3D results.\cite{KrMuCh05}
Experimentally, apart from pair tunneling,\cite{WaOrPh97,KoRavO06} the
bipolaron effect should also manifest itself in shot-noise
measurements.\cite{YuLi05,barthold:246804}  

In summary, we have presented unbiased quantum Monte Carlo results for
bipolaron formation in a two-dimensional quantum dot, 
taking into account the crystal lattice and quantum phonon
effects. Confinement is found to give rise to---respectively strengthen---a
bound state at intermediate to strong electron-phonon interaction, and to
reduce the critical coupling for bipolaron formation. The present
method can be used to study more general models with dispersive
phonons or more complicated electronic bands and dot
geometries.\cite{dRLa85,HoFe06}

\begin{acknowledgements}
M.H. gratefully acknowledges financial support by the Austrian Science Fund
(FWF) through the Erwin-Schr\"odinger Grant No.~J2583. We thank H.~Fehske and
V.~Heine for valuable discussion.
\end{acknowledgements}

%%%%%%%%%%%%%%%%%%%%%%%%%%%%%%%%%%%%%%%%%%%%%%%%%%%%%%%%%%%%%%%%%%%%%%%%%%%%%
%%%%%%%%%%%%%%%%%%%%%       BIBLIOGRAPHY              %%%%%%%%%%%%%%%%%%%%%%%
%%%%%%%%%%%%%%%%%%%%%%%%%%%%%%%%%%%%%%%%%%%%%%%%%%%%%%%%%%%%%%%%%%%%%%%%%%%%%

%\bibliography{../../17paper_iop_optcond/bibliography}
%\bibliographystyle{prsty}

\end{document}